\documentclass[aps,showpacs,twocolumn]{revtex4}
\usepackage{epsfig}
\usepackage{graphicx,color,dcolumn}
\usepackage{epstopdf}
\usepackage{amsmath,amssymb}
\usepackage{multirow}
\usepackage{diagbox}
\usepackage{changes}
\usepackage{booktabs}
\usepackage{threeparttable}
\usepackage{subfigure}
\usepackage{hyperref}

\begin{document}

\noindent\makebox[\textwidth]{\footnotesize Submitted to Chinese Physics C}
\vspace{10pt}

\title{ Investigation of Resonances in the $\Sigma({1/2}^{-})$ System Based on the Chiral Quark Model  }
\author{
Yu Yao\textsuperscript{a},
Xuejie Liu\textsuperscript{d},
Xiaoyun Chen\textsuperscript{c},
Yuheng Wu\textsuperscript{a},
Jialun Ping\textsuperscript{b},
Yue Tan\textsuperscript{a},
Qi Huang\textsuperscript{b}
}
\email[E-mail: ]{tanyue@ycit.edu.cn (Corresponding author) }
\email[E-mail: ]{huangqi@njnu.edu.cn (Corresponding author)}
\affiliation{\textsuperscript{a}Department of Physics, Yancheng Institute of Technology, Yancheng 224000, People's Republic of China }
\affiliation{\textsuperscript{b}Department of Physics, Nanjing Normal University, Nanjing 210023, People's Republic of China }
\affiliation{\textsuperscript{c}College of Science, Jinling Institute of Technology, Nanjing 211169, People's Republic of China }
\affiliation{\textsuperscript{d}Department of Physics, Henan Normal University, Xinxiang 453007, People's Republic of China }
\date{\today}

\begin{abstract}
In this work, we investigate the resonance structures in the $\Sigma(1/2^-)$ system from both three-quark and five-quark perspectives within the framework of the chiral quark model. An accurate few-body computational approach, the Gaussian Expansion Method, is employed to construct the orbital wave functions of multiquark states. To reduce the model dependence on parameters, we fit two sets of parameters to check the stability of the results. The calculations show that our results remain stable despite changes in the parameters. In the three-quark calculations, two $\Sigma(1/2^-)$ states are obtained with energies around 1.8~GeV, which are good candidates for the experimentally observed $\Sigma(1750)$ and $\Sigma(1900)$. In the five-quark configuration, several stable resonance states are identified, including $\Sigma \pi$, $N \bar{K}$, and $N \bar{K}^{*}$. These resonance states survive the channel-coupling calculations under the complex-scaling framework and manifest as stable structures. Our results support the existence of a two-pole structure for the $\Sigma(1/2^-)$ system, predominantly composed of $\Sigma \pi$ and $N \bar{K}$ configurations, analogous to the well-known $\Lambda(1380)$-$\Lambda(1405)$ ($\Sigma \pi$-$N \bar{K}$) system. On the other hand, although the energy of the $N \bar{K}^{*}$ configuration is close to that of $\Sigma(1750)$ and $\Sigma(1900)$, the obtained width is not consistent with the experimental values. This suggests that the $N \bar{K}^{*}$ state needs to mix with three-quark components to better explain the experimental $\Sigma(1750)$ and $\Sigma(1900)$ states. According to our decay width calculations, the predicted two resonance states are primarily composed of $\Sigma \pi$ and $N \bar{K}$, with their main decay channel being $\Lambda \pi$. Therefore, we encourage experimental groups to search for the predicted two-pole structure of the $\Sigma(1/2^-)$ system in the invariant mass spectrum of $\Lambda \pi$.
\end{abstract}

\maketitle

\section{Introduction}

The search for exotic states has long been a hot topic in hadron physics, with multiquark configurations playing a crucial role in their interpretation. In the framework of the traditional quark model, baryons are described as three-quark states and mesons as quark-antiquark pairs. This picture has been remarkably successful in describing a large number of ground-state hadrons\cite{Gell-Mann:1964ewy,Zweig:1964ruk}. However, it faces significant challenges in explaining certain excited hadrons, such as the $X(3872)$\cite{Tan:2019qwe} and $\Lambda(1405)$\cite{Tan:2025kjk}. It is now widely believed that many excited hadrons may have substantial multiquark components in their internal structure. Understanding the interactions among multiquark systems and revealing their underlying structure is essential for deepening our comprehension of quantum chromodynamics (QCD).

\begin{figure}[htp]
  \setlength {\abovecaptionskip} {-0.8cm}
  \centering
  \resizebox{0.50\textwidth}{!}{\includegraphics[width=5.5cm,height=5.5cm]{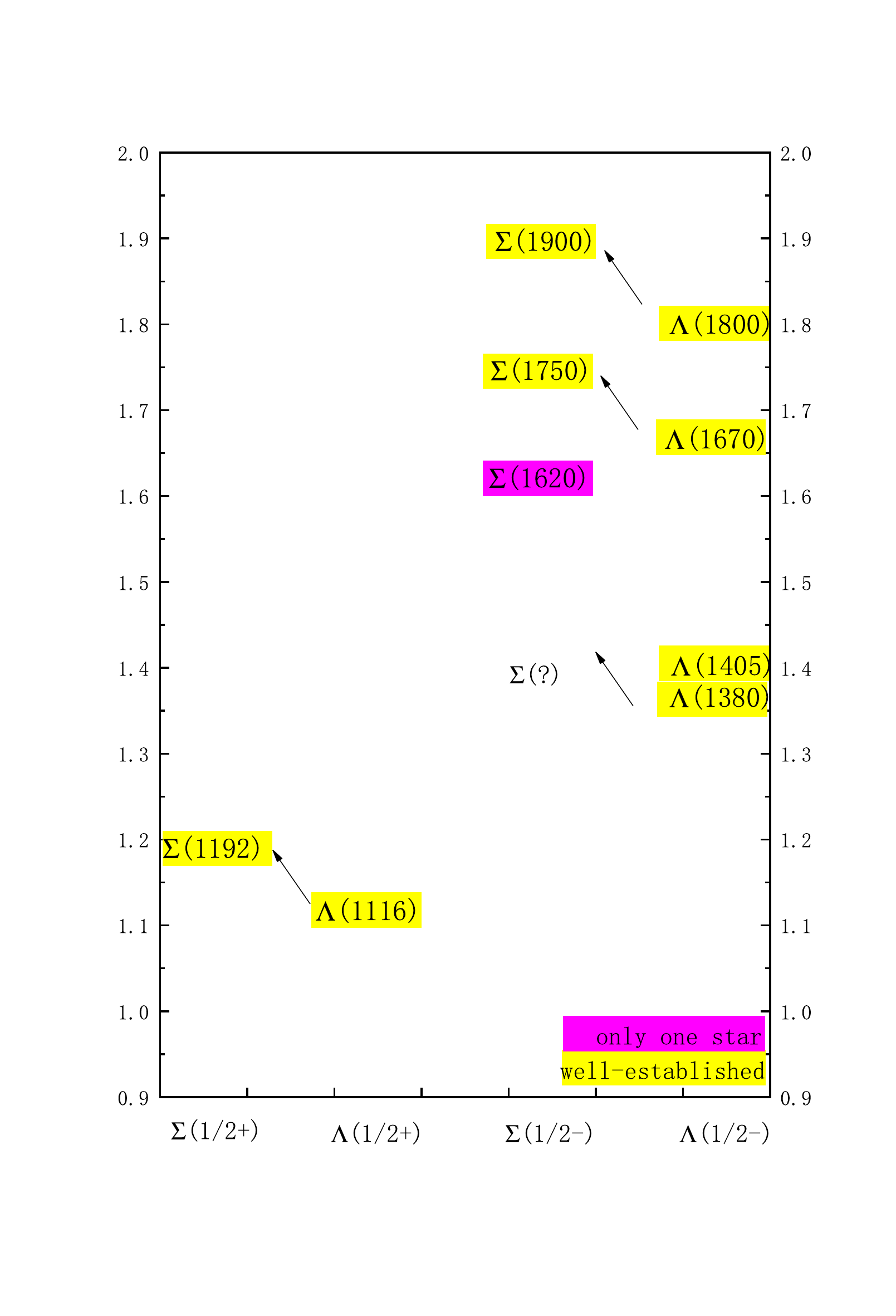}}
  \caption{Comparison of energy levels between $\Sigma$ and $\Lambda$ baryons.}
\label{Fic_SL}
\end{figure}

Among the excited baryons, the resonance structures in the $\Sigma(1/2^-)$ system have attracted sustained interest. Experimentally, four $\Sigma(1/2^-)$ states have been observed to date~\cite{ParticleDataGroup:2022pth}, namely $\Sigma(1620)$, $\Sigma(1750)$, $\Sigma(1900)$, and $\Sigma(2110)$. However, various theoretical studies ~\cite{Zhang:2004xt,Wu:2009nw,Wu:2009tu,Huang:2024oai,Wang:2024jyk,Wang:2024ewe} have suggested the possible existence of a stable structure in the mass range of $1.3$-$1.4$~GeV. In Ref.~\cite{Zhang:2004xt}, the authors constructed a diquark picture for the pentaquark with $J^{P} = \frac{1}{2}^{-}$ and predicted the lowest mass of the $\Sigma(1/2^{-})$ state to be around 1.36~GeV, while the corresponding lowest $\Lambda(1/2^{-})$ state lies at 1.44~GeV. This result is consistent with the observed $\Lambda(1405)$. Using the effective Lagrangian approach, Ref.~\cite{Wu:2009nw} investigated the $K^{-}p \rightarrow \Lambda \pi^{-} \pi^{+}$ reaction near the $\Lambda^{*}(1520)$ peak and found evidence for a new resonance around 1.38~GeV. This result was later confirmed in Ref.~\cite{Wu:2009tu} through a re-examination of older $K^{-}p \rightarrow \Lambda \pi^{-} \pi^{+}$ data, which further suggested that the width of this resonance could be larger than previously estimated. More recently, Ref.~\cite{Huang:2024oai} explored the role of triangle singularities in the process $J/\psi \rightarrow \Lambda \bar{\Lambda} \pi$. It was found that the triangle singularity plays a significant role in shaping the $\Lambda \pi$ invariant mass spectrum, leading to the appearance of a resonance-like structure around 1.4~GeV. Similarly, within the triangle singularity framework and using an effective Lagrangian approach, Ref.~\cite{Wang:2024ewe} investigated the resonance production in the reaction $\Lambda_c^+ \rightarrow \gamma \pi^+ \Lambda$, predicting a resonance near 1.38~GeV. It is worth noting that the $\Sigma$ baryons, as the isospin partners of the $\Lambda$ states, exhibit a similar spectral structure. As shown in Fig.~\ref{Fic_SL}, the energy of the $\Sigma(1/2^+)$ state is approximately 76~MeV higher than that of the $\Lambda(1/2^+)$ state. In the $J^P = 1/2^-$ sector, if we disregard the experimentally uncertain $\Sigma(1620)$ state (which currently has only a one-star rating), a parallel pattern can be observed: $\Sigma(1750)$ and $\Sigma(1900)$ correspond to $\Lambda(1670)$ and $\Lambda(1800)$, respectively. Based on this correspondence, the well-established $\Lambda(1405)$ and the tentative $\Lambda(1380)$ (though the latter also lacks strong experimental confirmation) would naturally imply the existence of a $\Sigma$ resonance in the mass range of 1.3-1.4~GeV.

 It is widely accepted that the $\Lambda(1405)$ and the tentative $\Lambda(1380)$ arise from a two-pole structure \cite{Mai:2014xna,Mai:2020ltx,Qin:2020gxr,Shevchenko:2011ce} generated by the coupled-channel interaction between the $N\bar{K}$ and $\Sigma\pi$ channels. That is, under strong channel coupling effects, both $N\bar{K}$ and $\Sigma\pi$ components remain dynamically stable. In this work, within the framework of the chiral quark model and with the help of a high-precision few-body method, the Gaussian Expansion Method (GEM), we first calculate the energy of the $\Sigma(1/2^-)$ system in the three-quark configuration. We then focus on the five-quark picture, where we perform bound-state calculations for the $\Sigma(1/2^-)$ system in various physical channels, including $N\bar{K}$, $\Sigma\pi$, and others. Subsequently, the complex-scaling method, a powerful tool for identifying resonance states, is employed to examine the stability of the resonances found in the bound-state calculations. In our analysis, the effects of channel coupling are thoroughly considered, as we couple the obtained states to all relevant scattering channels.

The structure of this paper is as follows. After the introduction, Sec.~II provides a brief description of the quark model, the construction of wave functions, and an overview of the complex-scaling method. Our numerical results and related discussions are presented in Sec.~III. Finally, a summary is given in Sec.~IV.
\section{Model setup} \label{wavefunction and chiral quark model}
\subsection{Chiral quark model}

In this work, the chiral quark model is employed to investigate the $\Sigma(1/2^-)$ system. The chiral quark model \cite{Vijande:2004he,Tan:2024omp,Tan:2024pqs} has become one of the most widely used approaches in hadron spectroscopy, hadron-hadron interactions, and the study of multiquark states, owing to its successful explanation of a large amount of experimental data. In this model, in addition to one-gluon exchange, the massive constituent quarks interact with each other via Goldstone boson exchange. Furthermore, color confinement and meson exchange are also incorporated into the model. The Hamiltonian of the chiral quark model is expressed as follows:
\begin{eqnarray}
H &=&\sum_{i=1}^n ( m_i +\frac{\vec{p}_{i}^2}{2 m_{i}} ) -T_{c} +  \sum_{i<j=1}^n V(r_{ij}) ,
\end{eqnarray}
where $m_i$ denotes the quark mass, $\vec{p}_i$ represents the quark momentum, $T_c$ is the center-of-mass kinetic energy of the quark system, and $V(r_{ij})$ represents the potential term. In the Jacobi coordinate system, for a three-quark system, the kinetic term $\sum_{i=1}^n \left( m_i + \frac{\vec{p}_{i}^2}{2 m_i} \right) - T_c$ can be reduced to
\begin{eqnarray}
\label{T3}
\frac{\vec{p}_{12}^2}{2\mu_{12}} + \frac{\vec{p}_{12,3}^2}{2\mu_{12,3}} ,
\end{eqnarray}

and for a five-quark system, it can be reduced to:

\begin{eqnarray}
\label{T5}
\frac{\vec{p}_{12}^2}{2\mu_{12}} + \frac{\vec{p}_{12,3}^2}{2\mu_{12,3}} + \frac{\vec{p}_{45}^2}{2\mu_{45}} + \frac{\vec{p}_{123,45}^2}{2\mu_{123,45}}.
\end{eqnarray}

Since we are studying light hadronic states (with quark constituents \( u \), \( d \), and \( s \)), the chiral potential, including the Goldstone boson exchanges (\(\pi\), \(\eta\), \(K\)) and scalar meson exchanges (\(a_0\), \(f_0\),  \(\kappa\), \(\sigma\)), plays a crucial role. Among these, the scalar meson exchanges are particularly important for resolving the \(\rho-\omega\) mass inversion problem \cite{Tan:2022pzi}. The interaction potentials are as follows:
\begin{eqnarray}
V_{\pi}(r_{ij}) &=& \frac{g_{ch}^2}{4\pi} \frac{m_{\pi}^2}{3m_im_j} \frac{\Lambda_{\pi}^2 m_\pi}{\Lambda_{\pi}^2 - m_{\pi}^2} \hat{S}_i \cdot \hat{S}_j \sum_{a=1}^3 \lambda_i^a \lambda_j^a \nonumber\\
&&\times \left[ Y(m_\pi r_{ij}) - \frac{\Lambda_{\pi}^3}{m_{\pi}^3} Y(\Lambda_{\pi} r_{ij}) \right], \nonumber \\
V_{K}(r_{ij}) &=& \frac{g_{ch}^2}{4\pi} \frac{m_{K}^2}{3m_im_j} \frac{\Lambda_K^2 m_K}{\Lambda_K^2 - m_K^2}  \hat{S}_i \cdot \hat{S}_j \sum_{a=1}^3 \lambda_i^a \lambda_j^a\nonumber\\
&&\times \left[ Y(m_K r_{ij}) - \frac{\Lambda_K^3}{m_K^3} Y(\Lambda_K r_{ij}) \right], \\
V_{\eta}(r_{ij}) &=& \frac{g_{ch}^2}{4\pi} \frac{m_{\eta}^2}{3m_im_j} \frac{\Lambda_{\eta}^2 m_{\eta}}{\Lambda_{\eta}^2 - m_{\eta}^2} \hat{S}_i \cdot \hat{S}_j \left( \lambda_i^8 \lambda_j^8 \cos\theta_P \right.\nonumber\\
&&\left. - \lambda_i^0 \lambda_j^0 \sin \theta_P \right)\left[ Y(m_\eta r_{ij}) - \frac{\Lambda_{\eta}^3}{m_{\eta}^3} Y(\Lambda_{\eta} r_{ij}) \right], \nonumber \\
V_{s}(r_{ij}) & = &  v_{\sigma}({{\bf r}_{ij}}) \lambda_i^0 \lambda_j^0+v_{a_0}({{\bf r}_{ij}})\sum_{a=1}^{3} \lambda_i^a \lambda_j^a \nonumber\\
&&v_{\kappa}({{\bf r}_{ij}})\sum_{a=4}^{7}
	\lambda_i^a \lambda_j^a+v_{f_0}({{\bf r}_{ij}}) \lambda_i^8 \lambda_j^8,{s=a_0,f_0, \kappa,\sigma} \nonumber\\
v_{s} & = & -\frac{g^2_{ch}}{4\pi} \frac{\Lambda^2_s}{\Lambda^2_s-m^2_s}m_s
	\left[ Y(m_{s}r_{ij})-\frac{\Lambda_s}{m_s}Y(\Lambda_{s}r_{ij})\right],   \nonumber\\ \nonumber
\end{eqnarray}
where, \( \boldsymbol{\lambda}^{a} \) represents the \(SU(3)\) Gell-Mann matrices that act on the flavor wave functions of the quark system. The Yukawa function \( Y(x) \) is explicitly defined as $Y(x) = \frac{e^{-x}}{x}$, where \( \Lambda_{\chi} \) is the cut-off parameter, and \( \frac{g_{ch}^2}{4\pi} \) corresponds to the coupling constant between the Goldstone bosons and the quarks. The masses of the Goldstone bosons, \( \pi \), \( K \), and \( \eta \), are denoted by \( m_{\pi} \), \( m_{\eta} \), and \( m_K \), respectively. Meanwhile, the mass of the scalar meson \( m_{\sigma} \) is related to the pion mass as $m_{\sigma}^2 \approx m_{\pi}^2 + 4 m_{u,d}^2.$

As for confinement potential, we utilize the following quadratic form in this study. It was demonstrated long ago by Goldman \cite{Goldman:1975dc} that in a relativistic first-order dynamical system, an interaction energy that increases linearly with the fermion separation has a broad range where a harmonic approximation is applicable for the second-order reduction of the equations of motion. Additionally, hadrons are typically small in size, and within this range, the difference between linear and quadratic confinement potentials is negligible. Furthermore, this difference can be effectively absorbed by the parameters \(a_c\) and \(\Delta\) in the quadratic potential,

\begin{align}
    V_{con}(r_{ij}) &= \left( -a_{c} r_{ij}^{2} - \Delta \right) \boldsymbol{\lambda}_i^c \cdot \boldsymbol{\lambda}_j^c.
\end{align}

The one-gluon exchange potential \(V_{oge}(r_{ij})\)  can be written as

\begin{eqnarray}
 V_{oge} (r_{ij}) &=& \frac{\alpha_s}{4} \boldsymbol{\lambda}_i^c \cdot \boldsymbol{\lambda}_j^c \left[\frac{1}{r_{ij}} - \frac{2}{3m_i m_j} \hat{S}_i \cdot \hat{S}_j\right. \nonumber\\
 && \times \left.\frac{e^{-r_{ij}/r_0(\mu_{ij})}}{r_{ij} r_0^2(\mu_{ij})} \right],
\end{eqnarray}
where \( \boldsymbol{\lambda}^c \) refers to the \(SU(3)\) Gell-Mann matrices acting on the color wave functions of the quark system, \(r_0\) is a model parameter, \(\alpha_s\) is the coupling constant, determined through experimental fitting, and \(\hat{S}_i\) represents the spin operator acting on the spin-\(\frac{1}{2}\) wave functions of the quarks.


After fitting the ground states of light mesons and baryons, all the model parameters are determined, which are collected into Table~\ref{modelparameters},
\begin{table}[tbp]
\begin{center}
\caption{Quark model parameters ($m_{\pi}=0.7$ $fm^{-1}$, $m_{\sigma}=3.42$ $fm^{-1}$, $m_{\eta}=2.77$ $fm^{-1}$, $m_{K}=2.51$ $fm^{-1}$). The fourth column corresponds to the first set of parameters, while the fifth column corresponds to the second set of parameters.\label{modelparameters}}
\begin{tabular}{cccc}
\hline\hline\noalign{\smallskip}
Quark masses       &$m_u=m_d$(MeV)               &490         &400\\
                   &$m_{s}$(MeV)                 &511         &550\\

Goldstone bosons
                   &$\Lambda_{\pi}(fm^{-1})$     &3.5         &3.5\\
                   &$\Lambda_{\eta}(fm^{-1})$    &2.2         &2.2\\
                   &$\Lambda_{\sigma}(fm^{-1})$  &7.0         &7.0\\
                   &$\Lambda_{a_0}(fm^{-1})$     &2.5         &2.5\\
                   &$\Lambda_{f_0}(fm^{-1})$     &1.2         &1.2\\
                   &$g_{ch}^2/(4\pi)$            &0.54        &0.54\\
                   &$\theta_p(^\circ)$           &-15         &-15\\

Confinement        &$a_{c}$ (MeV$\cdot fm^{-2}$) &98          &120\\
                   &$\Delta_{qq/q\bar{q}}$(MeV)  &-91.1/-10.1 &-92.4/-20.1\\
                   &$\Delta_{qs/q\bar{s}}$(MeV)  &-58.0/-10.0 &-60.4/-20.0\\
                   &$\Delta_{s\bar{s}}$(MeV)     &-18.1       &-18.1\\

OGE                & $\alpha_{qq/q\bar{q}}$      &0.69/1.34   &0.61/1.31 \\
                   & $\alpha_{qs/q\bar{s}}$      &0.90/1.15   &0.95/1.16\\
                   & $\alpha_{s\bar{s}}$         &0.91        &0.91\\
                   &$\hat{r}_0$(MeV)             &80.9        &85.1 \\
\hline\hline
\end{tabular}
\end{center}
\end{table}
while the fit results are presented in Table~\ref{fitresults}.

\begin{table}[t]
\begin{center}
\caption{Results of the hadron spectrum calculation. QM.1 and QM.2 represent the mass of the states under the two different sets of parameters.}
\label{fitresults}
\begin{tabular}{ccccccc}
\hline\hline\noalign{\smallskip}
~~~$IJ^{P}$~~~                &~~~ state~~~                    &~~~QM.1~~~&~~~QM.2~~~&~~~ PDG\cite{ParticleDataGroup:2022pth}   \\
$10^{-}$                      &  $\pi  $                       &  143   &  147   &  139 \\
$00^{-}$                      &  $\eta $                       &  599   &  623   &  547 \\
$11^{-}$                      &  $\rho $                       &  786   &  720   &  770 \\
$01^{-}$                      &  $\omega$                      &  800   &  733   &  782 \\
$\frac{1}{2}0^{-}$            &  $K$                           &  495   &  512   &  495 \\
$\frac{1}{2}1^{-}$            &  $K^{*}$                       &  915   &  893   &  892 \\
$00^{-}$                      &  $\eta^{\prime}$               &  804   &  828   &  957 \\
$01^{-}$                      &  $\phi$                        &  1029  &  1052  & 1020 \\
$\frac{1}{2}\frac{1}{2}^{+}$  &  $N$                           &  939   &  939   &  939 \\
$\frac{3}{2}\frac{1}{2}^{+}$  &  $\Delta$                      &  1271  &  1297  & 1232 \\
$0\frac{1}{2}^{+}$            &  $\Lambda$                     &  1071  &  1063  & 1116 \\
$1\frac{1}{2}^{+}$            &  $\Sigma$                      &  1215  &  1224  & 1226 \\
$1\frac{3}{2}^{+}$            &  $\Sigma^{*}$                  &  1345  &  1363  & 1384 \\
$1\frac{1}{2}^{+}$            &  $\Sigma(\frac{1}{2}^{-})$     &  1782  &  1826  & 1750 \\
$1\frac{1}{2}^{+}$            &  $\Sigma^{*}(\frac{1}{2}^{-})$ &  1829  &  1889  & 1900 \\
$\frac{1}{2}\frac{1}{2}^{+}$  &  $\Xi$                         &  1369  &  1363  & 1318 \\
$\frac{1}{2}\frac{1}{2}^{+}$  &  $\Xi^{*}$                     &  1479  &  1491  & 1553 \\
$\frac{3}{2}\frac{1}{2}^{+}$  &  $\Omega$                      &  1671  &  1694  & 1672 \\
\hline\hline
\end{tabular}
\end{center}
\end{table}
\subsection{The wave function of $\Sigma(1/2^{-})$ system}
Since our goal is to study the \( \Sigma(1/2^-) \) system from both the three-quark and five-quark perspectives, in this section, we present our constructions for the wave functions (\( \Psi^P \) and \( \Psi^B \)) of \( \Sigma(1/2^-) \). \( \Psi^B \) represents the three-quark structure wave function, while \( \Psi^P \) denotes the five-quark structure wave function. The total wave function \( \Psi \) is the tensor product of the color part \( \xi^l \), the flavor part \( \psi^k \), the orbital part \( \phi^i \), and the spin part \( \chi^j \).

\begin{eqnarray}
\Psi_{J,mJ}^{B_{ijkl}}(r) &=& \mathcal{A}_3 \left[ \phi_l^{B_i} \chi^{B_j} \right]_{J,mJ} \psi_{I,mI}^{B_k} \xi^{B_l}, \\
\Psi_{J,mJ}^{P_{ijkl}}(r) &=& \mathcal{A}_5 \left[ \phi_l^{P_i} \chi^{P_j} \right]_{J,mJ} \psi_{I,mI}^{P_k} \xi^{P_l},
\end{eqnarray}
where, \( \mathcal{A}_3 \) and \( \mathcal{A}_5 \) denote the antisymmetrization operators for the three-quark and five-quark wave functions, respectively.

For the orbital wave function \( \phi_{nlm} \) (which we will denote as \( \phi_l^i \) for convenience elsewhere in this paper), we adopt the GEM\cite{Hiyama:2003cu} for the expansion. In the GEM framework, the wave function with principal quantum number \( n \), orbital quantum number \( l \), and magnetic quantum number \( m \) can be expressed as

\begin{align}
\phi_{nlm}(r)  = N_{nl}r^{l} e^{-\nu_{n}r^2}Y_{lm}(r),
\end{align}
with $N_{nl}$ being the normalization constants as
\begin{align}
N_{nl}=\left[\frac{2^{l+2}(2\nu_{n})^{l+\frac{3}{2}}}{\sqrt{\pi}(2l+1)}
\right]^\frac{1}{2}.
\end{align}
For the three-quark system, there are two relative motions, whereas for the five-quark system, there are four relative motions. We denote the total orbital wave function for the three-quark system as \( \phi_{L,m_L}^{B}(r) \), and the total orbital wave function for the five-quark system as \( \phi_{L,m_L}^{P}(r) \). These wave functions can be written as

\begin{eqnarray}
\phi_{L,m_L}^{B_1}(r) &=& \phi_{l_{12}}(r_{12}) \phi_{l_3}(r_3), \\
\phi_{L,m_L}^{P_1}(r) &=& \phi_{l_{12}}(r_{12}) \phi_{l_3}(r_3) \phi_{l_{45}}(r_{45}) \phi_{l_{123,45}}(R).
\end{eqnarray}
The subscripts in the wave functions represent the relative motion between specific quarks. For example, the subscript \( 12 \) refers to the relative motion between quarks 1 and 2, and \( 123,45 \) refers to the relative motion between quarks 1, 2 and 3 with respect to quarks 4 and 5.

The spin wave function of the three-quark system can be of two possible spin configurations: \( S = \frac{1}{2} \) and \( S = \frac{3}{2} \). For \( S = \frac{1}{2} \), it can be obtained by coupling the spin \( S_1 = \frac{1}{2} \) of quark 1 with the spin \( S_2 = \frac{1}{2} \) of quark 2, resulting in the intermediate spin \( S_{12} = 0, 1 \), and then coupling this with the spin \( S_3 = \frac{1}{2} \) of the third quark to obtain the total spin \( S = \frac{1}{2} \). We denote the former case as \( \chi_{\frac{1}{2}}^{B_1} \) and the latter as \( \chi_{\frac{1}{2}}^{B_2} \). The \( S = \frac{3}{2} \) configuration is denoted as \( \chi_{\frac{3}{2}}^{B_3} \).
\begin{eqnarray}
\chi_{\frac{1}{2},\frac{1}{2}}^{B_1} &=& \frac{1}{\sqrt{2}} \left( \alpha \beta \alpha - \beta \alpha \alpha \right), \\
\chi_{\frac{1}{2},-\frac{1}{2}}^{B_1} &=& \frac{1}{\sqrt{2}} \left( \alpha \beta \beta - \beta \alpha \beta \right), \\
\chi_{\frac{1}{2},\frac{1}{2}}^{B_2} &=& \frac{1}{\sqrt{6}} \left( 2 \alpha \alpha \beta - \alpha \beta \alpha - \beta \alpha \alpha \right), \\
\chi_{\frac{1}{2},-\frac{1}{2}}^{B_2} &=& \frac{1}{\sqrt{6}} \left( \alpha \beta \beta + \beta \alpha \beta - 2 \beta \beta \alpha \right),\\
\chi_{\frac{3}{2},\frac{3}{2}}^{B_3} &=& \alpha \alpha \alpha, \\
\chi_{\frac{3}{2},\frac{1}{2}}^{B_3} &=& \frac{1}{\sqrt{3}} (\alpha \alpha \beta + \alpha \beta \alpha + \beta \alpha \alpha), \\
\chi_{\frac{3}{2},-\frac{1}{2}}^{B_3} &=& \frac{1}{\sqrt{3}} (\alpha \beta \beta + \beta \alpha \beta + \beta \beta \alpha), \\
\chi_{\frac{3}{2},-\frac{3}{2}}^{B_3} &=& \beta \beta \beta.
\end{eqnarray}

The spin wave function of the five-quark system can be considered as the coupling of the spin of a three-quark subsystem and the spin of a quark-antiquark pair. Given that \( J^P = \frac{1}{2}^- \), this can be obtained by coupling the following spin configurations: \( \frac{1}{2} \otimes 0 \), \( \frac{1}{2} \otimes 1 \), and \( \frac{3}{2} \otimes 1 \). According to the previous discussion, when the spin of the three-quark system is \( \frac{1}{2} \), there are two possible configurations. Therefore, we can obtain five total spin \( \frac{1}{2} \) wave functions. These are denoted sequentially as \( \chi_{\frac{1}{2}}^{P_1} \), \( \chi_{\frac{1}{2}}^{P_2} \), \( \chi_{\frac{1}{2}}^{P_3} \), \( \chi_{\frac{1}{2}}^{P_4} \), and \( \chi_{\frac{1}{2}}^{P_5} \).
\begin{eqnarray*}
\chi_{\frac{1}{2}, \frac{1}{2}}^{P_1} &=& \frac{1}{\sqrt{2}} (\alpha \beta \alpha- \beta \alpha \alpha )\times \frac{1}{\sqrt{2}} (\alpha \beta - \beta \alpha  ) , \\
\chi_{\frac{1}{2}, \frac{1}{2}}^{P_2} &=&  \frac{1}{\sqrt{6}} \left( 2 \alpha \alpha \beta - \alpha \beta \alpha - \beta \alpha \alpha \right)\times \frac{1}{\sqrt{2}} (\alpha \beta - \beta \alpha  ), \\
\chi_{\frac{1}{2}, \frac{1}{2}}^{P_3} &=& \frac{1}{\sqrt{12}}(\alpha \beta\alpha\alpha\beta+\alpha\beta\alpha\beta\alpha-\beta\alpha\alpha\alpha\beta-\beta\alpha\alpha\beta\alpha \\
&&-2\alpha\beta\beta\alpha\alpha-2\beta\alpha\beta\alpha\alpha  ), \\
\chi_{\frac{1}{2}, \frac{1}{2}}^{P_4} &=& \frac{1}{\sqrt{36}}(2\alpha\alpha\beta\alpha\beta+2\alpha\alpha\beta\beta\alpha-\alpha\beta\alpha\alpha\beta-\alpha\beta\alpha\beta\alpha\\
&&-\beta\alpha\alpha\alpha\beta-\beta\alpha\alpha\beta\alpha-2\alpha\beta\beta\alpha\alpha\\
&&+2\beta\alpha\beta\alpha\alpha-4\beta\beta\alpha\alpha\alpha  ),\\
\chi_{\frac{1}{2}, \frac{1}{2}}^{P_5} &=& \frac{1}{\sqrt{18}} ( \alpha\beta\beta\alpha\alpha+\beta\alpha\beta\alpha\alpha+\beta\beta\alpha\alpha\alpha-\alpha\alpha\beta\alpha\beta\\
&&-\alpha\alpha\beta\beta\alpha-\alpha\beta\alpha\alpha\beta-\alpha\beta\alpha\beta\alpha-\beta\alpha\alpha\alpha\beta\\
&&-\beta\alpha\alpha\beta\alpha+3\alpha\alpha\alpha\beta\beta).
\end{eqnarray*}

The isospin of the \( \Sigma(1/2^-) \) system is 1. In the three-quark framework, its flavor wave function \( \psi_{I=1}^{B_1} \) can be expressed as

\begin{eqnarray}
\psi_{I=1}^{B_1} = \frac{1}{\sqrt{2}}( uds + dus).
\end{eqnarray}

In the five-quark framework, it has two possible flavor combinations: \( qqq - \bar{q}s \) and \( qqs - \bar{q}q \). In the first flavor structure, \( qqq - \bar{q}s \), the total isospin \( I = 1 \) coupling scheme is \( \frac{1}{2} \otimes \frac{1}{2} \) and \( \frac{3}{2} \otimes \frac{1}{2} \). Since \( \frac{1}{2} \) can be obtained by coupling \( 0 \otimes \frac{1}{2} \) or \( 1 \otimes \frac{1}{2} \), we can obtain three total isospin wave functions. These are denoted as \( \psi_{I=1}^{P_1} \), \( \psi_{I=1}^{P_2} \), and \( \psi_{I=1}^{P_3} \).
\begin{eqnarray}
\psi_{I=1}^{P_1} &=& \frac{1}{\sqrt{2}}(udu\bar{d}s -duu\bar{d}s), \\
\psi_{I=1}^{P_2} &=& \frac{1}{\sqrt{6}}(2uud\bar{d}s -udu\bar{d}s-duu\bar{d}s), \\
\psi_{I=1}^{P_3} &=& \frac{1}{\sqrt{12}}(uud\bar{d}s +udu\bar{d}s+duu\bar{d}s+3 uuu\bar{u}s).
\end{eqnarray}
In the second flavor structure, \( qqs - \bar{q}q \), the total isospin \( I = 1 \) coupling schemes are \( 0 \otimes 1 \), \( 1 \otimes 0 \), and \( 1 \otimes 1 \). These are denoted as \( \psi_{I=1}^{P_4} \), \( \psi_{I=1}^{P_5} \), and \( \psi_{I=1}^{P_6} \).
\begin{eqnarray}
\psi_{I=1}^{P_4} &=& \frac{1}{\sqrt{2}}(uds\bar{d}u -dus\bar{d}u), \\
\psi_{I=1}^{P_5} &=& \frac{1}{\sqrt{2}}(uus\bar{u}u +uus\bar{d}d), \\
\psi_{I=1}^{P_6} &=& \frac{1}{\sqrt{4}}(-uds\bar{d}u - dus\bar{d}u - uus\bar{d}u+ uus\bar{d}d).
\end{eqnarray}

For the color wave function in a three-quark system, it must be in a color-neutral state. Hence, it is expressed as:
\begin{eqnarray}
\xi^{B} &=& \frac{1}{\sqrt{6}} \left( {rgb} - {grb} + {gbr} - {brg} + {bgr} \right).
\end{eqnarray}
In the case of a five-quark system, when considering a molecular state configuration, the color wave function can be viewed as the combination of the three-quark and two-quark color wave functions. This can be obtained by coupling two color singlet states, which is represented by \( \xi^{P_1} \), and written as:
\begin{eqnarray}
\xi^{P_1} &=& \frac{1}{\sqrt{6}} \left({rgb} - {rbg} + {gbr} - {grb} + {brg} -{bgr} \right) \nonumber \\
&& \times \frac{1}{\sqrt{3}} \left( \bar{r}r + \bar{g}g + \bar{b}b \right).
\end{eqnarray}

\subsection{Complex-Scaling Method}
The complex-scaling method, introduced in Refs. \cite{Aguilar:1971ve,Balslev:1971vb}, is a robust technique for identifying resonant states. This method involves replacing the spatial coordinates \( \vec{r} \) in the Hamiltonian \( H \) with \( \vec{r} e^{i\theta} \), where \( \theta \) is a complex scaling factor. By solving the Schrodinger equation in the complex plane, both the energy and decay width (lifetime) of the resonance can be simultaneously determined.
\begin{figure}[htbp]
    \centering
    \includegraphics[width=0.5\textwidth, trim=4.5cm 5.0cm 4.5cm 2.0cm, clip]{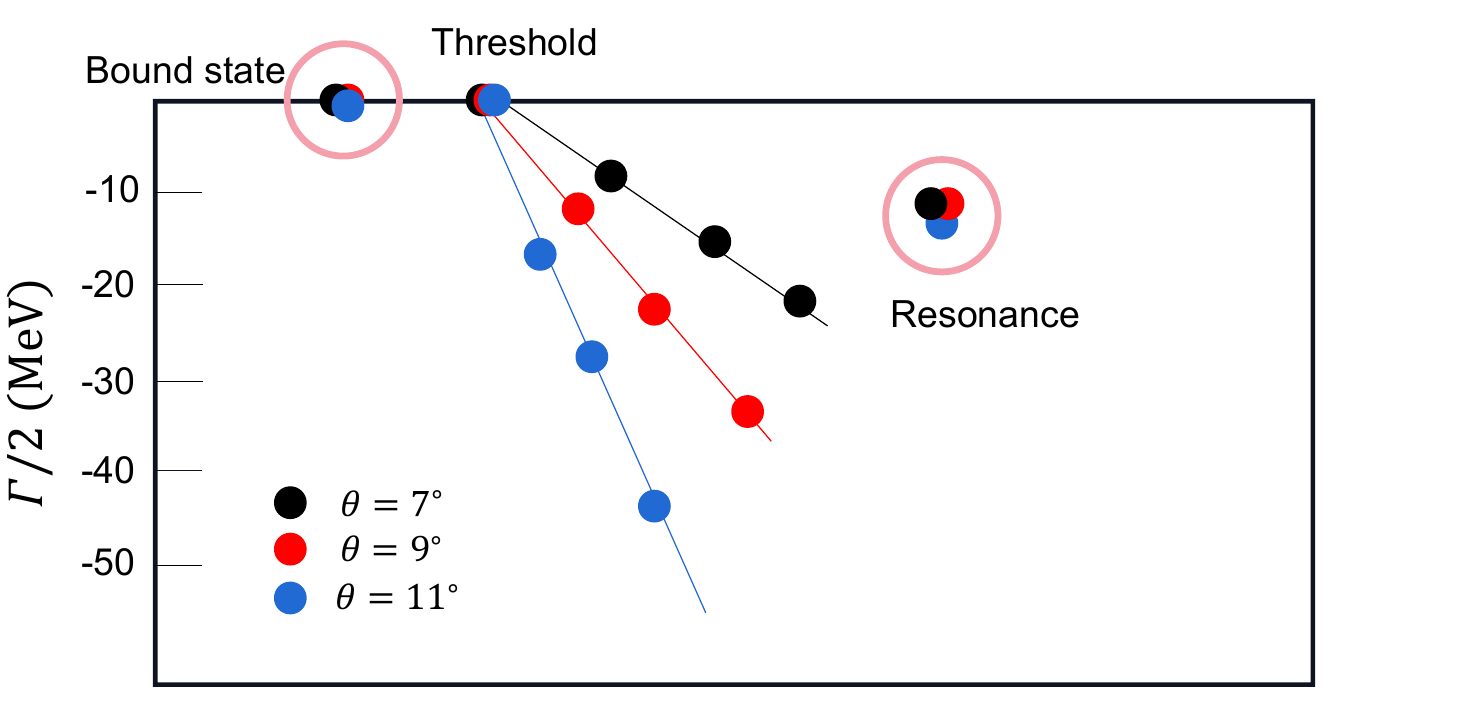}
    \caption{Schematic complex energy distribution}
     \label{csexample}
  \end{figure}

In the complex-scaling framework, the real part of the complex energy, \( M \), is plotted along the horizontal axis, while the half-width, \( \Gamma/2 \), is plotted along the vertical axis. As the scaling angle \( \theta \) varies, the system shows different characteristic behaviors. The method allows for the simultaneous study of bound states, resonances, and scattering states, which are shown in Fig. \ref{csexample}. Their behaviors as \( \theta \) changes are as follows:

\begin{itemize}
    \item \textbf{Bound states:} For bound states, the corresponding points converge to the real energy axis. The position of these points on the real axis directly gives the mass of the bound state.
    \item \textbf{Scattering states:} Points corresponding to scattering states lie along the scattering line, where \( \theta \) is constant. Since the scattering spectrum is continuous, an infinite number of points would theoretically lie along this line. To avoid clutter, only a few representative points are shown to illustrate the general trend within the energy range.
    \item \textbf{Resonances:} Resonance points do not lie on the scattering continuum (the scattering line) but stay fixed as \( \theta \) changes. The vertical position of these points represents the half-width, \( \Gamma/2 \). It is important to note that the resonance width is a result of coupling to open scattering channels, and hence, no width is predicted for purely three-quark or two-quark systems.
\end{itemize}

A key advantage of the complex-scaling method is that transforming the system into the complex coordinate plane significantly enhances the analysis of resonant states. This makes it a powerful tool for exploring resonance phenomena in systems governed by strong interactions.

\section{Results and Discussions}

In this section, we first explore the possible internal quark configurations of the $\Sigma(1/2^{-})$ resonance within the three-quark framework. Then, we carry out bound-state calculations within the five-quark model to identify the physical channels exhibiting attractive interactions, which may indicate the formation of resonance states. Within the complex-scaling framework, these resonance candidates are coupled to all relevant scattering channels to distinguish genuine resonances from unstable ones. To assess the stability of the computed results, all calculations are performed under two different parameter sets. Additionally, to estimate the decay widths of stable resonances and analyze the scattering behavior of unstable ones, we further carry out channel-coupling calculations between each resonance candidate and the corresponding open channels.
\subsection{Three-Quark Calculation}

The baryon energies obtained from the three-quark calculations are listed in Table~\ref{fitresults}. As shown, under the two parameter sets, the ground states $\Sigma(1/2^+)$ and $\Sigma(3/2^+)$ are in good agreement with experimental observations. Based on this, under the first set of parameters, their respective first excited states are found at 1782~MeV and 1829~MeV, while under the second set of parameters, their respective first excited states are found at 1826~MeV and 1889~MeV, which are consistent with the experimentally observed $\Sigma(1750)$ and $\Sigma(1900)$, respectively. From another perspective, the mass differences between the ground states $\Sigma(1/2^+)$, $\Sigma(3/2^+)$ and their excited counterparts $\Sigma(1750)$, $\Sigma(1900)$ are both around 500~MeV. This is comparable to the average mass difference between the $1P$ and $1S$ states of mesons such as $\rho(770)$-$a_1(1260)$ (490 MeV), $K^{*}(895)$-$K_1(1270)$ (375 MeV), $K(495)$-$K_1(1270)$ (775 MeV), $\eta^{\prime}(958)$-$f_1(1420)$ (462 MeV), and $\phi(1020)$-$f_1(1420)$ (400 MeV). The average of these values is approximately 500 MeV.

Therefore, it is reasonable to consider that the experimentally observed $\Sigma(1750)$ and $\Sigma(1900)$ are likely dominated by three-quark components, similar to the $\Lambda(1670)$ and $\Lambda(1800)$ resonances in the $\Lambda(1/2^-)$ sector. If there exists a one-to-one correspondence between the $\Sigma(1/2^-)$ and $\Lambda(1/2^-)$ families, then  $\Sigma(1/2^-)$ resonances should exist in the 1.3-1.4~GeV mass region, analogous to the $\Lambda(1405)$ and $\Lambda(1380)$, and predominantly composed of five-quark components.

\subsection{Five-Quark Calculations}

\subsubsection{Bound-State Calculation}
\begin{table*}[tp]
  \centering
  \fontsize{9}{8}\selectfont
  \makebox[\textwidth][c]{
   \begin{threeparttable}
   \caption{\label{S1400} The energies of the $\Sigma^{*}(\frac{1}{2}^{-})$  system. $i,j,k,l$ stands for the index of orbit, flavor, spin and color wave functions, respectively. $E_{th}$ means the threshold of corresponding channel, $E_{sc}$ is the energy of every single channel, $E_{mix}$ is the lowest energy of the system by coupling all channels. (unit: MeV)}
    \begin{tabular}{ccccccccccc}
\hline\hline
$\Psi_{\frac{1}{2},\frac{1}{2}}^{P_{ijkl}}$ ~~~~ &Channel~~~~&~~~~$E_{th1}$~~~~ &~~~~$E_{sc1}$~~~~&~~~~$B.E.$~~~~&$E_{mix1}$&~~~~$E_{th2}$ ~~~~~&~~~~$E_{sc2}$~~~&~~~~$B.E.$~~~&~~~$E_{mix2}$~~~   \\ \hline
$\Psi_{\frac{1}{2},\frac{1}{2}}^{P_{1111}}$/$\Psi_{\frac{1}{2},\frac{1}{2}}^{P_{1261}}$&$N \bar{K}$  &1435.3   &1417.4  & 17.9&1215.6 &1452.6    &1443.1&  9.5&1214.0         \\
$\Psi_{\frac{1}{2},\frac{1}{2}}^{P_{1221}}$/$\Psi_{\frac{1}{2},\frac{1}{2}}^{P_{1461}}$&$N \bar{K}^*$&1854.3   &1839.7  & 14.6&       &1834.2    &1826.4&  7.8&                         \\
$\Psi_{\frac{1}{2},\frac{1}{2}}^{P_{1351}}$~~~  &$\Delta \bar{K}^{*}$                                &2186.4   &2188.4  &   ub&       &2191.1    &2193.5&   ub&                         \\
$\Psi_{\frac{1}{2},\frac{1}{2}}^{P_{1621}}$~~~  &$\Sigma \pi$                                        &1358.5   &1348.5  & 10.0&       &1372.8    &1366.6&  6.2&                         \\
$\Psi_{\frac{1}{2},\frac{1}{2}}^{P_{1641}}$~~~  &$\Sigma \rho$                                       &2001.8   &1995.1  &  6.7&       &1958.6    &1957.9&  0.7&                         \\
$\Psi_{\frac{1}{2},\frac{1}{2}}^{P_{1651}}$~~~  &$\Sigma^* \rho$                                     &2132.6   &2130.3  &  2.3&       &2096.9    &2097.8&   ub&                         \\
$\Psi_{\frac{1}{2},\frac{1}{2}}^{P_{1411}}$~~~  &$\Lambda \pi$                                       &1214.9   &1216.8  &   ub&       &1211.5    &1214.2&   ub&                         \\
$\Psi_{\frac{1}{2},\frac{1}{2}}^{P_{1431}}$~~~  &$\Lambda \rho$                                      &1858.2   &1859.5  &   ub&       &1797.3    &1799.6&   ub&                         \\
$\Psi_{\frac{1}{2},\frac{1}{2}}^{P_{1521}}$~~~  &$\Sigma \eta$                                       &1814.8   &1816.2  &   ub&       &1848.8    &1851.3&   ub&                         \\
$\Psi_{\frac{1}{2},\frac{1}{2}}^{P_{1541}}$~~~  &$\Sigma \omega$                                     &2015.3   &2016.1  &   ub&       &1945.8    &1948.0&   ub&                         \\
$\Psi_{\frac{1}{2},\frac{1}{2}}^{P_{1551}}$~~~  &$\Sigma^*\omega$                                    &2146.1   &2144.9  &  1.2&       &2084.8    &2085.3&   ub&                         \\
\hline\hline
    \end{tabular}
   \end{threeparttable}}
  \end{table*}

The results for the five-quark system are listed in Table \ref{S1400}. The $\Sigma(1/2^{-})$ five-quark system has two possible quark configurations: $qqq$-$\bar{q}s$ and $qqs$-$\bar{q}q$. In the first quark configuration, $qqq$-$\bar{q}s$, the total isospin $I = 1$ coupling occurs in the forms $\frac{1}{2} \otimes \frac{1}{2}$ and $\frac{3}{2} \otimes \frac{1}{2}$. For the total spin $J = \frac{1}{2}$, the coupling schemes are $\frac{1}{2} \otimes 0$, $\frac{1}{2} \otimes 1$, and $\frac{3}{2} \otimes 1$. The quantum combinations corresponding to these are $N\bar{K}$, $N\bar{K}^{*}$, and $\Delta \bar{K}^{*}$.  In the second quark configuration, $qqs$-$\bar{q}q$, the total isospin $I = 1$ coupling can be $0 \otimes 1$, $1 \otimes 0$, or $1 \otimes 1$, while the total spin $J = \frac{1}{2}$ coupling includes $\frac{1}{2} \otimes 0$, $\frac{1}{2} \otimes 1$, and $\frac{3}{2} \otimes 1$. The corresponding quantum combinations for these are $\Lambda \pi$, $\Lambda \rho$, $\Sigma \pi$, $\Sigma \rho$, $\Sigma \eta$, $\Sigma \omega$, $\Sigma^{*} \omega$, and $\Sigma^{*} \rho$.


Therefore, in our calculations, we considered a total of 11 physical channels. Under the first set of parameters, the energy distribution of these channels spans the range from 1.2 GeV to 2.1 GeV. Among them, we identified 6 bound states, which are $N\bar{K}$, $N\bar{K}^{*}$, $\Sigma \pi$, $\Sigma \rho$, $\Sigma^{*} \omega$, and $\Sigma^{*} \rho$. The binding energies of the first three channels, $N\bar{K}$, $N\bar{K}^{*}$, and $\Sigma \pi$, are all greater than 10 MeV, while the binding energies of the remaining three channels are only a few MeV. This indicates that the first three channels, namely $N\bar{K}$, $N\bar{K}^{*}$, and $\Sigma \pi$, are more likely to form stable resonant states. Under the second set of parameters, these states, $N\bar{K}$, $N\bar{K}^{*}$, and $\Sigma \pi$, remain stable and are still bound states, with binding energies around $6\sim9$ MeV. In addition, $\Sigma \rho$ has a binding energy of 6.7 MeV under the first set of parameters, but only 0.7 MeV under the second set, indicating that it is not very stable. Therefore, under both parameter sets, $N\bar{K}$, $N\bar{K}^{*}$, $\Sigma \rho$, and $\Sigma \pi$ are bound states, while under the first set of parameters, $\Sigma^{*} \omega$ and $\Sigma^{*} \rho$, though also shallow bound states with binding energies of $1\sim2$ MeV, do not survive under the second set of parameters.

To understand this discrepancy, we systematically studied the contributions of each Hamiltonian term to the binding energy (threshold minus calculated value) for these six states, and the results are listed in Table~\ref{potential}. We observe that the kinetic energy predominantly contributes repulsive forces, while the attraction mainly comes from the $\sigma$-meson exchange, one-gluon-exchange, and $a_0$-meson exchange, with the $\sigma$-meson exchange playing a dominant role. The kinetic and potential energies compete with each other; as the attractive potential increases, the quarks are brought closer together, leading to an increase in the repulsive kinetic energy. Compared to the second set, the binding energies under the first set are generally larger, resulting in larger kinetic energy and stronger attractive potential contributions. We note that, under the second set of parameters, the $\sigma$-meson exchange for the $\Sigma^{*} \omega$ and $\Sigma^{*} \rho$ states provides insufficient attraction (less than 10 MeV), which is unable to overcome the repulsive kinetic energy and thus prevents these states from forming bound states.

\begin{table*}[htb]
\caption{\label{potential} The contributions of all potentials to the binding energy (unit: MeV) in $\Sigma(1/2^{-})$ five-quark system, where Q.M.1 and Q.M.2 represent the calculation results under two different sets of parameters.}
\begin{tabular}{cccccccccccccc}\hline\hline
& & ~~~~kinetic~~~~ & ~~~~con~~~~~ & ~~~~~oge~~~~~ & ~~~~~$\pi$~~~~~ &  ~~~~~$\eta$~~~~~ & ~~~~~$\sigma$~~~~~ & ~~~~~$a_0$~~~~~ & ~~~~$f_0$~~~~ \\
\multirow{2}{*}{$N\bar{K}$}~~~        & Q.M.1 & $90.0$ & $-5.6$ & $-12.0$ & $-2.7$ & $-1.1$ & $-78.1$ & $-6.6$ & $-3.0$  \\
                                      & Q.M.2 & $64.2$ & $-4.6$ & $ -6.2$ & $-2.1$ & $-0.0$  & $-54.0$ & $-4.6$ & $-2.3$   \\
\multirow{2}{*}{$N\bar{K}^{*}$}~~~    & Q.M.1 & $76.9$ & $-6.1$ & $ -8.3$ & $-2.1$ & $-0.1$ & $-66.7$ & $-5.7$ & $-2.7$  \\
                                      & Q.M.2 & $55.5$ & $-4.5$ & $ -4.2$ & $-1.7$ & $-0.0$  & $-46.7$ & $-4.0$ & $-2.1$   \\
\multirow{2}{*}{$\Sigma\pi$}~~~       & Q.M.1 & $64.9$ & $-4.0$ & $ -7.2$ & $2.5$ & $-0.1$ & $-62.5$ & $-3.6$ & $-0.0$  \\
                                      & Q.M.2 & $31.1$ & $-1.8$ & $ -3.3$ & $1.7$ & $-0.0$  & $-28.8$ & $-1.7$ & $-0.0$   \\
\multirow{2}{*}{$\Sigma\rho$}~~~      & Q.M.1 & $50.3$ & $-3.5$ & $ -5.2$ & $3.5$ & $-0.1$ & $-48.1$ & $-3.5$ & $-0.1$  \\
                                      & Q.M.2 & $20.1$ & $-1.3$ & $ -1.5$ & $2.9$ & $-0.0$  & $-18.5$ & $-1.1$ & $-0.0$   \\
\multirow{2}{*}{$\Sigma^{*}\rho$}~~~  & Q.M.1 & $27.6$ & $-2.2$ & $ -2.5$ & $ 0.0$ & $ 0.3$ & $-28.0$ & $-0.0$ & $-0.0$  \\
                                      & Q.M.2 & $9.6$ & $-0.6$ & $ -0.5$ &  $ 1.0$ & $0.0$  & $-8.7$ & $-0.0$ & $-0.0$   \\
\multirow{2}{*}{$\Sigma^{*}\omega$}~~~& Q.M.1 & $20.5$ & $-1.6$ & $ -0.3$ & $ 1.1$ & $0.2$ & $-21.0$ & $0.1$ & $0.0$  \\
                                      & Q.M.2 & $8.0$ & $-0.5$ & $ -0.1$ & $ 0.1$ & $0.0$  & $-7.0$ & $ 0.0$ & $0.0$   \\
\hline\hline
\end{tabular}
\end{table*}

\subsubsection{Resonance-State Calculation}

\begin{figure}[htp]
  \setlength {\abovecaptionskip} {-0.3cm}
  \centering
  \resizebox{0.5\textwidth}{!}{\includegraphics[width=11.5cm,height=7.0cm]{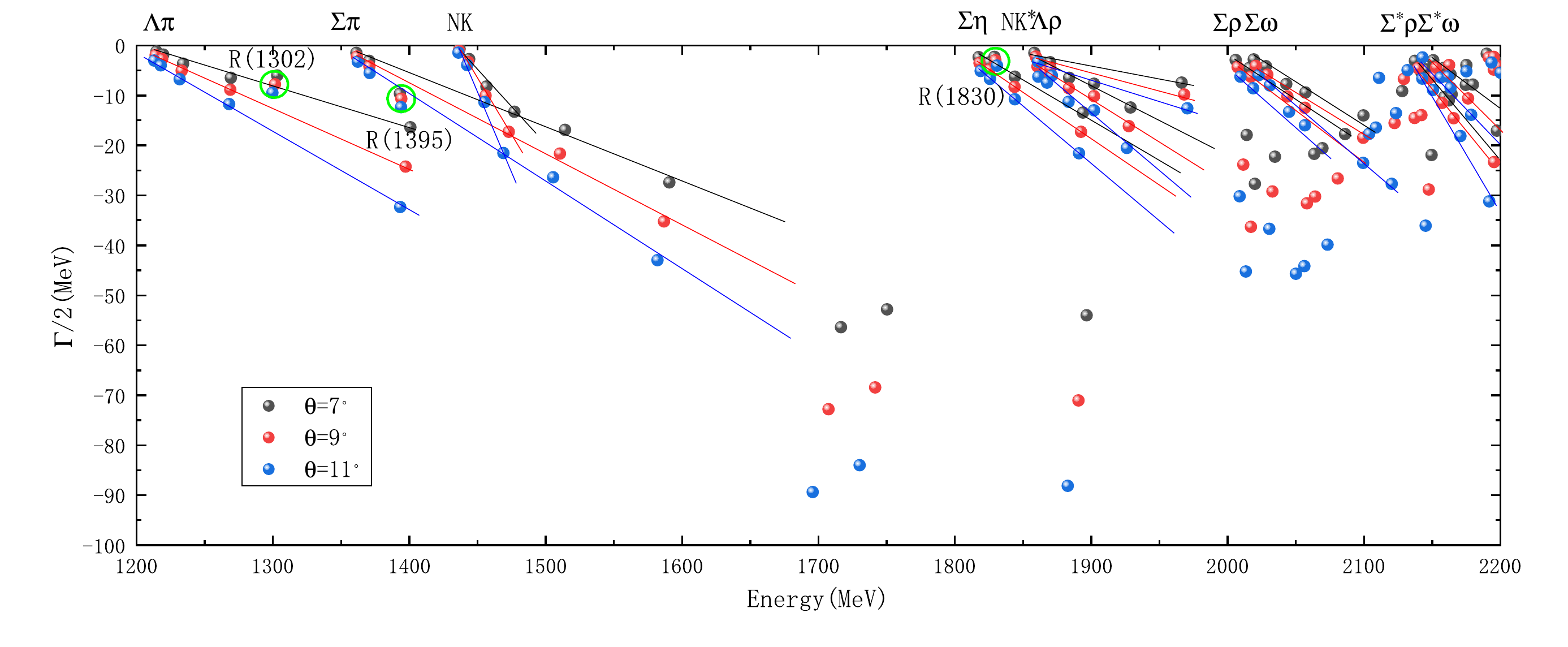}}
  \caption{Complex-scaling results for the $\Sigma(1/2^{-})$ five-quark system in the 1200-2200 MeV range for the first set of parameters.}
\label{Fic_All}
\end{figure}

\begin{figure}[htp]
  \setlength {\abovecaptionskip} {-0.3cm}
  \centering
  \resizebox{0.5\textwidth}{!}{\includegraphics[width=11.5cm,height=7.0cm]{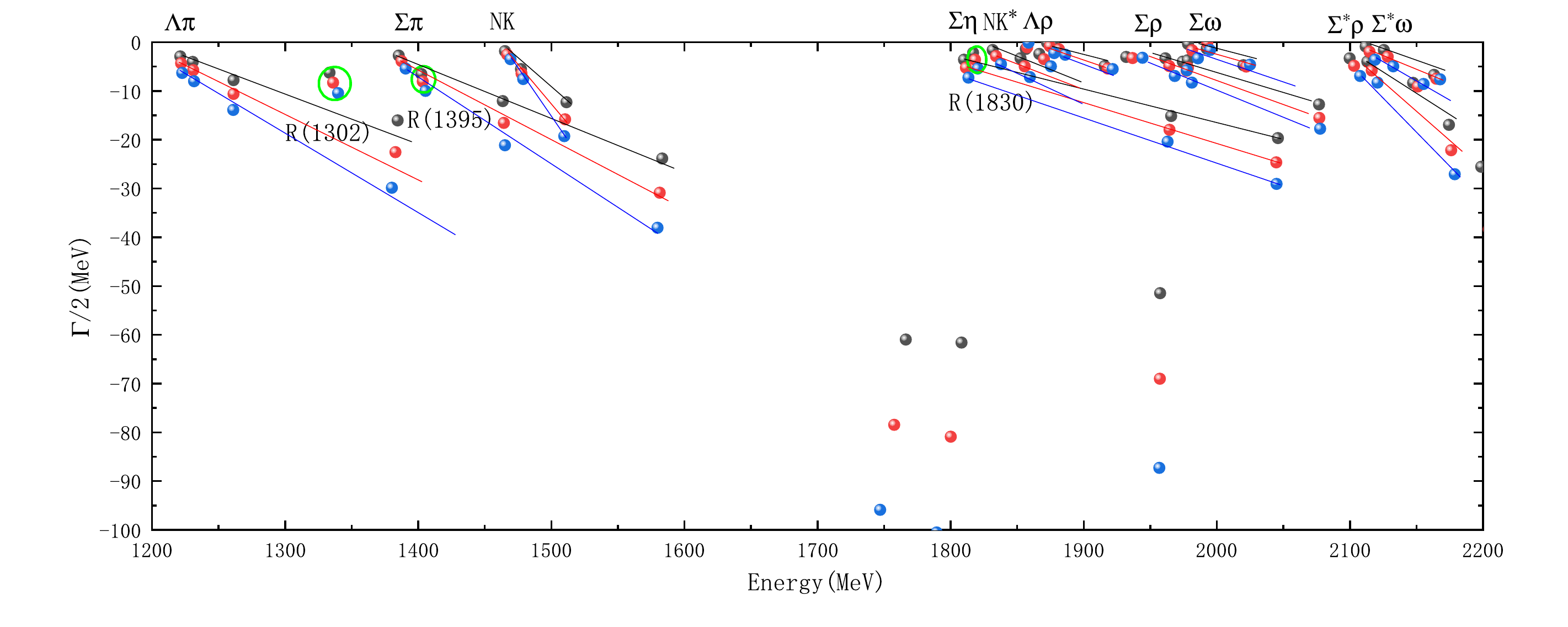}}
  \caption{Complex-scaling results for the $\Sigma(1/2^{-})$ five-quark system in the 1200-2200 MeV rangefor the second set of parameters.}
\label{Fic_All2}
\end{figure}

In this subsection, our goal is to use the effects of channel coupling to check whether the resonance candidates obtained from the bound state calculation survive in the channel coupling framework. Based on the bound state calculations under the two parameter sets, we identified three relatively stable bound states, namely $N\bar{K}$, $N\bar{K}^{*}$, and $\Sigma \pi$, one less stable bound state, $\Sigma \rho$, and two more controversial bound states, $\Sigma^{*} \omega$ and $\Sigma^{*} \rho$. Under both parameter sets, we first perform a complete channel coupling calculation using the complex-scaling method for all channels to search for stable resonance states. Then, we perform channel coupling calculations between the bound states identified in the previous calculations and the corresponding threshold channels to obtain the decay widths of the resonance states and investigate the channel coupling mechanisms (i.e., the reasons why unstable bound states transform into scattering states). The energy range of the five-quark system under investigation is between 1.2 GeV and 2.1 GeV. The purpose of the resonance-state calculation is to verify whether the resonance candidates identified in the bound state calculation correspond to genuine resonances. Therefore, in Figs. \ref{Fic_All} and \ref{Fic_All2}, we present the energy range from 1.2 GeV to 2.2 GeV.


As shown in Fig. \ref{Fic_All}, we obtained three stable resonance states, namely $R(1302)$, $R(1395)$, and $R(1830)$ (we use "R(Energy)" to denote Resonance.). We performed a component analysis of these three resonance states, and the results (see Table~\ref{percent}) show that $R(1302)$ and $R(1395)$ mainly consist of the $N \bar{K}$-$\Sigma\pi$ coupling. Specifically, $R(1302)$ has a dominant $\Sigma\pi$ component (about $53.1\%$), while $R(1395)$ has a dominant $N \bar{K}$ component (about $73.3\%$). On the other hand, $R(1830)$ has a main component of $N \bar{K}^{*}$ (about $51.6\%$) and an important $\Sigma\eta$ component (about $33.4\%$). This indicates that the $N \bar{K}^{*}$-$\Sigma\eta$ coupling is also significant, but there is no attraction between $\Sigma\eta$. Therefore, even though the $N \bar{K}^{*}$-$\Sigma\eta$ coupling is strong, it still cannot form a resonance state dominated by $\Sigma\eta$. Thus, their main components are $\Sigma \pi$, $N\bar{K}$, and $N\bar{K}^{*}$, respectively. Meanwhile, another shallow bound state $\Sigma^{*} \rho$, and two controversial bound states, $\Sigma^{*} \omega$ and $\Sigma^{*} \rho$, obtained in the previous bound-state calculations, did not survive in the channel coupling calculation and became scattering states. This conclusion is corroborated by the results under the second set of parameters, as shown in Fig. \ref{Fic_All2}, where $R(1302)$, $R(1395)$, and $R(1830)$ are still stable resonance states.
In our calculations, the energies of the predicted $R(1302)$ and $R(1395)$ are very close to each other, which corresponds to the two-pole structure observed experimentally in the $\Lambda(1/2^{-})$ system, namely $\Lambda(1380)$ and $\Lambda(1405)$. Since the $\Lambda(1/2^{-})$ system and the $\Sigma(1/2^{-})$ system only differ by isospin, the prediction of $R(1302)$ and $R(1395)$ is highly credible. The energy of $R(1830)$ lies between the $\Sigma(1750)$ and $\Sigma(1900)$, suggesting that $R(1830)$ is one of the five-quark candidates for these states.

\begin{table}[tp]
\centering
\caption{Various decay channels and corresponding decay widths of the obtained resonances. (unit: MeV)\label{Width}}
\begin{tabular}{cccccccccc}
 \hline    \hline
   Decay channels & $R(1302)$&  P    &$R(1395)$& P  &$R(1830)$  & P \\
\hline
$\Lambda \pi $       &  5.0      &100$\%$&  24.0   &74$\%$&  7.6  &37$\%$         \\
$\Sigma \pi $        &  ...      &...    &  8.6    &26$\%$&  4.2  &20$\%$        \\
$N \bar{K} $         &  ...      &...    &  ...    &...   &  4.1  &19$\%$         \\
$\Sigma \eta $       &  ...      &...    &  ...    &...   &  5.0  &24$\%$         \\
Total                &  5.0      &       &  32.6   &      &  20.9          \\
\hline \hline
\end{tabular}
\end{table}

\begin{table*}[htb]
\caption{\label{percent} Main components of the three stable resonance states.}
\begin{tabular}{cccccccccccccc}\hline\hline
         &$N \bar{K}$&$N \bar{K}^{*}$ &$\Delta\bar{K}^{*}$&$\Sigma\pi$&$\Sigma\rho$&$\Sigma^{*}\rho$&$\Lambda\pi$&$\Lambda\rho$&$\Sigma\eta$&$\Sigma\omega$&$\Sigma^{*}\omega$ \\
$R(1302)$&15.5$\%$   &0.1$\%$         &0.0$\%$            &53.1$\%$   &0.1$\%$     &0.0$\%$         &31.0$\%$    &0.1$\%$      &0.1$\%$     &0.0$\%$      &0.0$\%$   \\
$R(1395)$&73.3$\%$   &0.1$\%$         &0.0$\%$            &26.1$\%$   &0.1$\%$     &0.0$\%$         & 0.9$\%$    &0.1$\%$      &0.0$\%$     &0.0$\%$      &0.0$\%$   \\
$R(1830)$&1.5$\%$    &51.6$\%$        &0.0$\%$            &3.9$\%$   &0.1$\%$     &0.0$\%$         &4.5$\%$    &5.1$\%$      &33.4$\%$    &0.0$\%$      &0.0$\%$   \\
\hline\hline
\end{tabular}
\end{table*}

\begin{figure*}[htbp]  
  \centering
  \setlength{\abovecaptionskip}{-0.1cm}
  \resizebox{\textwidth}{!}{  
    \includegraphics[width=12cm, height=7cm]{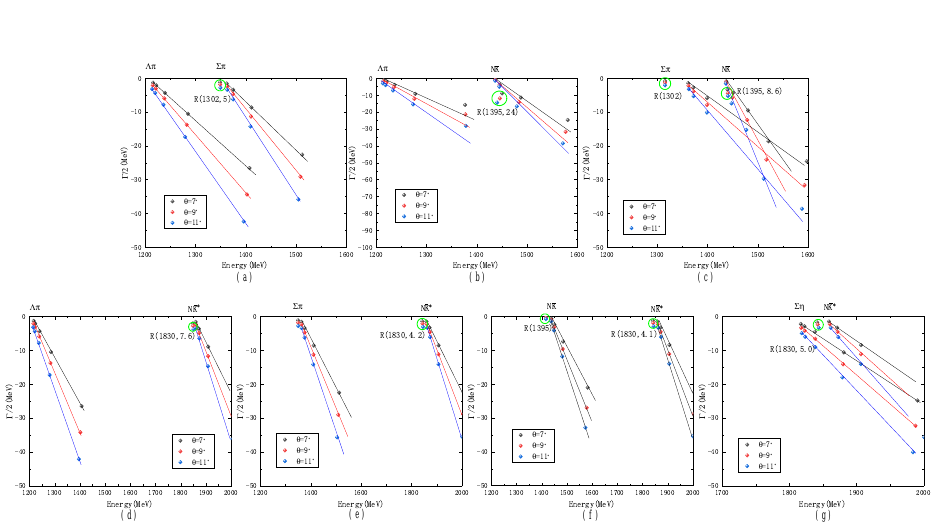} 
  }
  \caption{Calculated decay widths of resonance states in the $\Sigma(1/2^{-})$ five-quark system.(a)$\Sigma \pi \rightarrow \Lambda \pi$. (b)$N \bar{K} \rightarrow \Lambda \pi$. (c)$N \bar{K} \rightarrow \Sigma \pi$. (d)$N \bar{K}^{*} \rightarrow \Lambda \pi$. (e)$N \bar{K}^{*} \rightarrow \Sigma \pi$. (f)$N \bar{K}^{*} \rightarrow N \bar{K}$. (g)$N \bar{K}^{*} \rightarrow \Sigma \eta$.}
  \label{Fic_AR}
\end{figure*}

\begin{figure*}[htbp]  
  \centering
  \setlength{\abovecaptionskip}{-0.1cm}
  \resizebox{\textwidth}{!}{  
    \includegraphics[width=12cm, height=15.0cm]{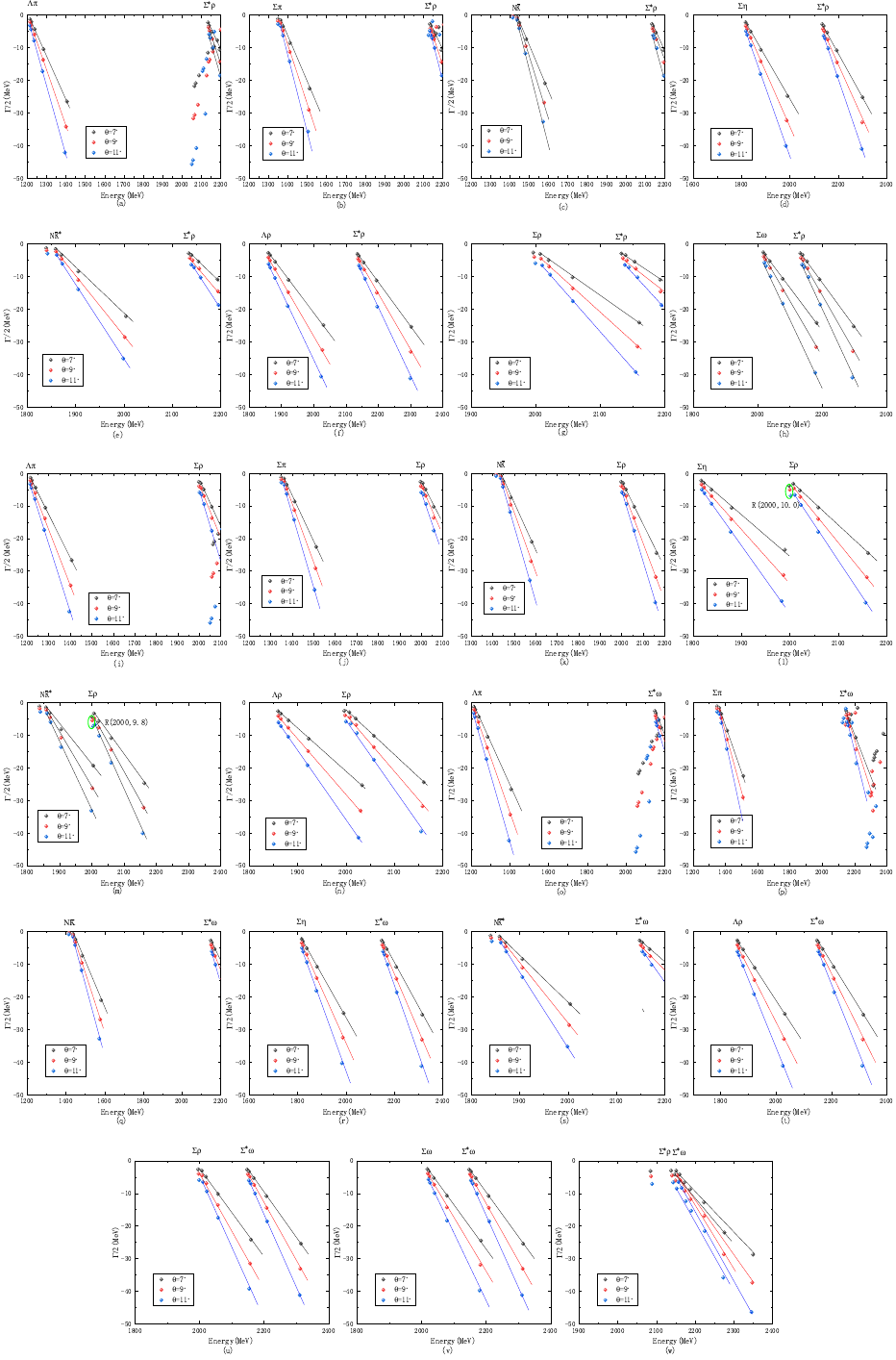} 
  }
\caption{ Channel-coupling effects transforming some unstable resonances into scattering states in the $\Sigma(1/2^{-})$ five-quark system. (a)$\Sigma^{*} \rho \rightarrow \Lambda \pi$.(b)$\Sigma^{*} \rho \rightarrow \Sigma \pi$. (c)$\Sigma^{*} \rho \rightarrow N\bar{K}$. (d)$\Sigma^{*} \rho \rightarrow \Sigma \eta$. (e)$\Sigma^{*} \rho \rightarrow N\bar{K}^{*}$. (f)$\Sigma^{*} \rho \rightarrow \Lambda \rho$. (g)$\Sigma^{*} \rho \rightarrow \Sigma \rho$. (h)$\Sigma^{*} \rho \rightarrow \Sigma \omega$. (i)$\Sigma \rho \rightarrow \Lambda \pi$. (j)$\Sigma \rho \rightarrow \Sigma \pi$. (k)$\Sigma \rho \rightarrow N\bar{K}$. (l)$\Sigma \rho \rightarrow \Sigma \eta$. (m)$\Sigma \rho \rightarrow N\bar{K}^{*}$. (n)$\Sigma \rho \rightarrow \Lambda \rho$.  (o)$\Sigma^{*} \omega \rightarrow \Lambda \pi$. (p)$\Sigma^{*} \omega \rightarrow \Sigma \pi$. (q)$\Sigma^{*} \omega \rightarrow N\bar{K}$. (r)$\Sigma^{*} \omega \rightarrow \Sigma \eta$. (s)$\Sigma^{*} \omega \rightarrow N\bar{K}^{*}$. (t)$\Sigma^{*} \omega \rightarrow \Lambda \rho$. (u)$\Sigma^{*} \omega \rightarrow \Sigma \rho$. (v)$\Sigma^{*} \omega \rightarrow \Sigma \omega$.(w)$\Sigma^{*} \omega \rightarrow \Sigma^{*} \rho$.}\label{Fic_AS}
\end{figure*}

To obtain the decay widths for the genuine resonance states and to investigate why the unstable resonances, such as $\Sigma \rho$, $\Sigma^{*} \omega$, and $\Sigma^{*} \rho$, decay into scattering states, we performed channel coupling calculations with all possible decay channels. The decay widths are shown in Fig. \ref{Fic_AR} and listed in Table \ref{Width}. The results of the unstable resonances coupling with the decay channels and transitioning into scattering states are shown in Fig. \ref{Fic_AS}. Since the resonance states obtained from both parameter sets are consistent in our calculations, the decay widths are only provided for the first parameter set. According to our calculations, in Fig. \ref{Fic_AR}, the decay width of $R(1302)$ to $\Lambda \pi$ is 5.0 MeV, and since this is the only decay channel, the total width is also 5.0 MeV. For $R(1395)$, the decay widths to $\Lambda \pi$ and $\Sigma \pi$ are 24.0 MeV and 8.6 MeV, respectively, giving a total width of 32.6 MeV. For $R(1830)$, the decay widths to $\Lambda \pi$, $\Sigma \pi$, $N\bar{K}$, and $\Sigma \eta$ are 7.6 MeV, 4.2 MeV, 4.1 MeV, and 5.0 MeV, respectively, resulting in a total width of 20.9 MeV.  Considering that the experimental widths of $\Sigma(1750)$ and $\Sigma(1900)$ are both over 100 MeV, from the perspective of decay widths, $R(1830)$ might require additional three-quark components to better explain the experimental $\Sigma(1750)$ and $\Sigma(1900)$ states, since its width in our calculation is only 20.9 MeV. Thus, we propose that the experimental $\Sigma(1750)$ and $\Sigma(1900)$ are likely to be mixed states, containing components of both the three-quark $\Sigma(1/2^{-})$ state and the five-quark $N\bar{K}^{*}$ state. Finally, based on the percentage of their decay widths, as shown in Table \ref{Width}, $R(1302)$ and $R(1395)$ predominantly decay into $\Lambda \pi$ (with more than 70\%), while $R(1830)$ has a more complex decay pattern, with both $\Lambda \pi$ and $\Sigma \eta$ playing significant roles.  According to Fig.~\ref{Fic_AS}, it can be seen that the $\Sigma \rho$ state survives only when coupled with the $\Sigma \eta$ and $N \bar{K}^{*}$ channels, and it rapidly decays into the $\Lambda \pi$, $\Sigma \pi$, $N \bar{K}$, and $\Lambda \rho$ channels. In contrast, the two more controversial resonance candidates, $\Sigma^{*} \omega$ and $\Sigma^{*} \rho$, do not survive in any of the coupled decay channels. The reason why $\Sigma \rho$, $\Sigma^{*} \omega$, and $\Sigma^{*} \rho$ fail to form genuine resonance states is that their binding energies are too small, leading them to transition into scattering states once coupled with most of the decay channels.

\section{Summary}

Within the framework of the chiral quark model, we have systematically studied the $\Sigma(1/2^-)$ system from both the three-quark and five-quark perspectives under two different parameter sets by employing the Gaussian Expansion Method (GEM).

From the three-quark calculations, we have obtained two $\Sigma(1/2^-)$ states with energies of approximately 1.75~GeV and 1.82~GeV for the first parameter set, and 1.83~GeV and 1.89~GeV for the second parameter set. These results are in good agreement with the experimentally observed $\Sigma(1750)$ and $\Sigma(1900)$, respectively, while simultaneously reproducing the ground states $\Sigma(1/2^+)$ and $\Sigma(3/2^+)$ with good accuracy. In addition, the energy gap between these excited and ground states is roughly 500~MeV, which is consistent with the typical mass difference between the $1P$ and $1S$ states in light meson systems such as $\eta$, $\rho$, $\omega$, $K$, and $K^*$. This observation further supports the interpretation that $\Sigma(1750)$ and $\Sigma(1900)$ possess substantial three-quark components.

In the bound-state calculations within the five-quark framework, we have obtained three stable bound states: $\Sigma \pi$, $N \bar{K}$, and $N \bar{K}^{*}$, along with a shallower bound state, $\Sigma \rho$, and two controversial bound states, $\Sigma^{*} \rho$ and $\Sigma^{*} \omega$. Our calculations show that, under the influence of channel coupling effects, the three stable bound states, $\Sigma \pi$, $N \bar{K}$, and $N \bar{K}^{*}$, form stable resonance states $R(1302)$, $R(1395)$, and $R(1830)$, while $\Sigma^{*} \rho$, $\Sigma^{*} \omega$, and $\Sigma \rho$ all transition into scattering states. Among these, $R(1302)$ and $R(1395)$, with their main components being the $\Sigma \pi$ and $N \bar{K}$ channels, despite being close in energy, survive the coupling process and remain stable, giving rise to a two-pole structure. These predicted two-pole resonances are similar in nature to the well-established $\Lambda(1380)$-$\Lambda(1405)$ system.  The main component of $R(1830)$ is $N \bar{K}^{*}$, which shows good agreement with the masses of $\Sigma(1750)$ and $\Sigma(1900)$. However, its decay width deviates from experimental data. This suggests that $R(1830)$ requires additional three-quark components to better explain the experimental $\Sigma(1750)$ and $\Sigma(1900)$ states. In future work, we intend to explore this system using an unquenched quark model to further test our hypothesis.

In summary, we obtained three resonance states in the $\Sigma(1/2^-)$ system: $R(1302)$, $R(1395)$, and $R(1830)$. Among these, $R(1302)$-$R(1395)$ exhibits a two-pole structure similar to the $\Lambda(1380)$-$\Lambda(1405)$ system, and we strongly recommend experimental efforts to search for these states in the invariant mass spectrum of $\Lambda \pi$. Although the mass of $R(1830)$ agrees well with the experimental $\Sigma(1750)$ and $\Sigma(1900)$, the calculated width is relatively small. This suggests that three-quark mixing effects may be needed to better explain the experimental $\Sigma(1750)$ and $\Sigma(1900)$ states.

\acknowledgments{This work is supported partly by the National Science Foundation of China under Contract No. 12205249 and No. 12305087. Y. T. is supported by the Funding for School-Level Research Projects of Yancheng Institute of Technology under Grant No. xjr2022039. Q. H. is supported by the Start-up Funds of Nanjing Normal University under Grant No. 184080H201B20.}

\end{document}